\documentclass[12pt,aps,preprint,nofootinbib,amsmath,amssymb]{revtex4}
\usepackage{graphicx}
\usepackage{color}
\usepackage{amsmath}
\begin{document}

\title{Quantum Mechanics of Individual Systems \footnote {This paper was published in {\sl American Journal of Physics} {\bf 36}, 704-712, (1967).  It has been reset and posted to arXiv to make it more accessible. It is otherwise unchanged and no attempt has been made to update the references. It is still useful for its derivation of the frequencies of outcomes in an infinite ensemble of identical systems described in Section III and the Mathematical Appendix. For example it has been employed in  derivations of Born's rule.  The paper  is also sometimes cited in expositions of  Quantum Bayesianism. \vskip .1in
Not surprisingly, in the more than 50 years since this paper was published, the author's views on quantum mechanics have expanded and to some extent changed. These  developments were not made necessary by errors in this paper. Rather they were necessary to  extend quantum mechanics as theory of the outcomes of  laboratory measurements to a quantum theory of a closed systems like the universe as a whole.  That is, it was necessary to formulate a quantum mechanics for cosmology in which the familiar quantum theory used here is an  approximation to the more general theory that is appropriate for describing measurement situations.  For some of these later developments see for instance  J.B. ~Hartle {\it The Quantum Mechanics of Cosmology}, in {\sl Quantum
Cosmology and Baby Universes:  Proceedings of the 1989 Jerusalem Winter
School for Theoretical Physics}, 
eds.~S. Coleman, J.B. Hartle, T. Piran, and S. Weinberg, 
WorldScientific, Singapore (1991) pp. 65-157, arXiv:1805.12246,
 and Murray Gell-Mann and James Hartle, {\it  Quantum Mechanics in the Light of Quantum Cosmology} in 
{\sl Complexity, Entropy, and the Physics of Information}, Santa Fe Institute
Studies in the Sciences of Complexity VIII, edited by W.H. Zurek, Addison-Wesley,
Reading, MA (1990), arXiv:1803.04605 .\vskip .2in}}

\author{J.B.~Hartle}
\affiliation{Department of Physics, University of California\\
Santa Barbara, CA 93106-9530}
\email{hartle@physics.ucsb.edu}

\date{April 1968}

\begin{abstract}
A formulation of quantum mechanics, which begins by postulating assertions for individual 
physical systems, is given. The statistical predictions of quantum mechanics for infinite 
ensembles are then derived from its assertions for individual systems. A discussion of the 
meaning of the ``state'' of an individual quantum mechanical system is given, and an 
application is made to the clarification of some of the paradoxical features of the theory.
\end{abstract}

\maketitle

\section{Introduction}

Conventional exegeses of quantum mechanics commonly begin with the fundamental
assertions that (1) the state of an individual physical system can be characterized by a
vector in Hilbert space and that (2) to every physically realizable value of a
measurable quantity there corresponds a subspace of this Hilbert space. It is further
asserted that the empirical content of these assumptions is that the absolute square of
the scalar product of the state vector with its projection on one of these subspaces
gives the {\it probability} that a measurement of the associated physical quantity will have
the value characteristic of the subspace as a result.      

In physics the assertion of a probability has an empirical significance, not for
an individual system in general, but only for an ensemble of identical systems. In the
latter case, the assertion of a probability of the result of a measurement can be
translated into an assertion about the relative frequencies of the results of many
identical experiments performed on identical members of the ensemble. In many branches
of classical physics, the empirical assertions about an ensemble of systems do not enter
the theory in a fundamental way, but are deducible from the assertions which can be made
about its members. This is not the case in the conventional formulation of quantum
mechanics described above. There, because the probability interpretation of the wave
function is a fundamental assumption, the predictions of the theory for ensembles of
identical systems are not deduced from its predictions for the individual systems. It is
not necessary, however, to formulate quantum mechanics in this way.

In this article we give an exposition of quantum mechanics from the point of view of the
individual system. We begin with a careful definition of what is meant by the words ``the
state of a system'', proceed to a discussion of how the formalism of quantum mechanics
describes individual systems, and finally derive the statistical assertions of quantum
mechanics from its assertions for individual systems. One of the main advantages of
this procedure is that it leads to a precise understanding of the character of a physical
state in quantum mechanics. This understanding leads to the resolution of many of the
seemingly paradoxical features of quantum mechanics although not in a way which will
carry the same degree of emotional appeal for all persons.

The purpose of this review is not to present a new theory of the foundation of quantum
mechanics. Certainly many of the ideas adumbrated here should have a familiar ring to
readers of the literature on this subject.\footnote{Some useful papers in which other
references are found are: (a) A.~Daneri, A.~Loinger, and G.R.~Prosperi, {\sl Nucl.~Phys.}
{\bf 33}, 297 (1962); (b) B. d'Espagnat, {\sl Conceptions de la physique contemporaine}
(Hermann et Cie., Paris, 1965); (c) J.M.~Jauch, {\sl Foundations of Quantum Mechanics}
(Addison-Wesley Pub.~Co., Reading, MA, 1968); (d) J.M.~Jauch, {\sl Hely.~Phys.~Acta} {\bf
37}, 193 (1964); (e) J.M.~Jauch, E.P.~Wigner, and M.M.~Yanase, {\sl Nuovo Cimento} {\bf
48}, 144 (1967); (f) H.~Margenau, {\sl  Phil.~Sci.} {\bf 30}, 1 (1963), {\sl ibid.} {\bf
30}, 138 (1963); (g) E.P.~Wigner, {\sl Am.~J.~Phys.} {\bf 31}, 6 (1963); (h) E.P.~Wigner, 
{\it Remarks on the Mind-Body Question} in {\sl The Scientist Speculates}, ed.~by I.J.~Good, 
(Basic Books, New, York, 1962.} Neither is the purpose of this paper to judge how
adequately quantum mechanics conforms to any previously conceived philosophically,
historically, or aesthetically-based standards for physical theory. The aim is to discuss
precisely the empirical significance of the terms involved in formulating a
quantum-mechanical theory (in particular the notion of state) in a way which, at least to
the author, leads clearly to an understanding of the theory uncomplicated by the
conventional paradoxes.

\section{The State of an Individual System}

Often the result of a measurement on an individual physical system is expressed by a
number. For the following analysis, however, it will be useful to note that every
measurement can be resolved into a number of experiments called propositions whose result
is either that they are true or false.\footnote{See, for example, G.~Birkhoff and J.~von
Neumann, {\sl Ann. Math} {\bf 37}, 835 (1936); G.W.~Mackey, {\sl Mathematical Foundations 
of Quantum Mechanics} (W.A.~Benjamin, Inc., New York, 1963); J.M.~Jauch, {\sl Foundations 
of Quantum Mechanics} (Addison-Wesley Pub.~Co., Reading, MA, 1968).} By way of example,
one may take a measurement of the angular momentum of an atom which may be resolved into
the propositions: ``The angular momentum is 0'', ``The angular momentum is 2h'', etc. For
every physical system the set of all propositions exhausts all possible measurements.

In classical physics, the state of an individual system is known when it is known whether
each proposition is true or false. {\it The state of an individual system in classical physics
is defined as the list of all possible propositions together with their truth values --- true
or false}. If a system is asserted to be in a certain state, what is being predicted is
the result of all and any of the measurements corresponding to the propositions. The
prediction can be verified or falsified by performing the measurements and checking
whether the experimental truth values agree with those in the predicted list.

Implicit in this classical definition of state is the assumption that all propositions
{\it can} be measured together, so that together they all can be asserted to be true or false.
Whether this assumption itself is correct is an empirical question to be determined by an
analysis of the measuring process. In fact, it is known that the assumption is not true.
The empirical evidence for this is the experimental verification of quantum mechanics
which predicts that noncommuting observables cannot be simultaneously specified and that 
therefore, by inference, they cannot be simultaneously measured.\footnote{We are
preserving here the fiction that to every Hermitian operator there corresponds an
experiment to observe the corresponding quantity. In fact, experimental arrangements are
known for only a few of these quantities. This remarkable circumstance seems to arise
from the fact that we are able to conceive of experimental arrangements only in
classical terms. The description of a quantum mechanical state, however, requires many
more numbers than a corresponding classical state (for example, the state of a spin-2
object in classical physics is described by two numbers while in quantum mechanics it is
described by nine), and there is a corresponding greater number of ``measurable''
quantities. The statements presented here do not depend on the assumption that every
operator corresponds to an observable quantity, but lacking a precise criterion for which
ones do correspond we present the results as if the assumption were true.} 

In quantum mechanics it is not possible to assert the truth values of all propositions at
once. An individual system cannot be prepared in such a way that all propositions are
true or false. Some propositions must necessarily be indefinite. {\sl The state of an
individual system in quantum mechanics is, therefore, defined as the list of all
propositions together with their truth values --- true, false, or indefinite.} A knowledge of
the state of a quantum-mechanical system permits the prediction of the results of only a
very limited class of experiments. These are the experiments which have the truth
values --- {\it true} or {\it false}, a situation which is abbreviated by saying that the 
corresponding proposition is {\it definite}. About propositions labeled {\it indefinite}, no 
predictions at all can be made for a single experiment on an individual system.

A quantum-mechanical state can be specified by listing all the true propositions. The
value {\it false} is then assigned to the negation of these and the value {\it indefinite} 
to the remaining ones.

 Not every assignment of truth values corresponds to a valid quantum-mechanical state. It 
is a remarkable assertion of quantum mechanics that for any physical system the set of
propositions is isomorphic to the set of subspaces of some Hilbert space. Since the state
of a system can be labeled by the true propositions, the states of a system can also be
put into correspondence with certain subspaces of the Hilbert space.\footnote{We ignore superselection rules here.} It is a further assertion of quantum mechanics that the
subspaces corresponding to states in which a maximum number of propositions are definite
are all one dimensional. These states are called pure, and to every pure state there
corresponds a ray in the Hilbert space. This statement is equivalent to one maintaining
that the propositions which correspond to different one-dimensional subspaces cannot be
simultaneously true.

A ray may, in turn, be labeled by one of its vectors. One then concludes that to every
pure state there corresponds a vector in a Hilbert space characteristic of the physical
system. The vector is a complete specification of the state because a knowledge of it
enables one to determine the truth value of any proposition. A proposition is true if the
state vector is an eigenfunction of the corresponding operator with eigenvalue $+1$ and 
false if it is an eigenfunction with eigenvalue 0. All other propositions are indefinite.
In general, a quantity is definite if the vector is an eigenfunction of the corresponding
operator and the definite value is the eigenvalue. Quantities for which the vector is not
an eigenfunction of the corresponding operator are indefinite.

From the point of view taken here, the state vector is simply a convenient shorthand for
the list of all measurable propositions and their truth values. The laws of quantum
mechanics determine how the state vector and this list evolve in time. A calculation of
the state vector at a future time enables predictions to be made of the results of
measurements on a single system of definite quantities and no predictions for the
indefinite ones. The state vector is then a summary of that information gained from past
manipulations of the individual system that can be used for making predictions about
which quantities are definite at later times.\footnote{The role of the wave function as a
summary of the past information relevant for future predictions is stated especially
clearly by E.P.~Wigner. See Ref. 1(g), 1(h).}

\section{Statistical Predictions of Quantum Mechanics Deduced from the Quantum Mechanics
of Individual Systems}

No mention of probability was made in the discussion of quantum mechanics for individual 
systems in the previous section. This is because the assertion of a probability has an
empirical significance not for any individual experiment but only as a statement about
the frequencies of results of ensembles of experiments. In this section we show how the
probability interpretation of the wave function results from an application of the
previous discussion to ensembles of identical systems themselves considered as individual
systems.

For simplicity, individual systems are considered those whose states are labeled by the
discrete eigenvalues $a_i$  of a single observable $A$
\begin{equation}
A | i\rangle = a_i|i\rangle\, .
\label{one}
\end{equation}
A general pure state\footnote{The following discussion can be generalized to include mixtures in a
straightforward way.} of a system can be denoted  by $|s\rangle$ and will always be taken
to be normalized to unity, $\langle s | s\rangle = 1$. Suppose that we have $N$ of these
systems, each identically prepared in a state $|s\rangle$. We distinguish the various
individual systems by using a label $\alpha, | s, \alpha\rangle, \alpha=1, \cdots, N$. 
The state vector for this ensemble of $N$ identically prepared systems is then
\begin{equation}
|(s)^N\rangle = |s, 1\rangle\otimes|s, 2\rangle\otimes | s, 3\rangle\otimes \cdots
\otimes | s, N\rangle\, .
\label{two}
\end{equation}
An infinite ensemble of identically prepared systems in the state $|s\rangle$ is described 
by the infinite product
\begin{equation}
|(s)^\infty\rangle = |s, 1\rangle \otimes |s, 2\rangle \otimes |s, 3\rangle \otimes
\cdots\, .
\label{three}
\end{equation}
The precise mathematical meaning of such products is given in the Appendix.

The result of a measurement of the quantity $A$ cannot be predicted for any individual system of
these ensembles. The quantity $A$ is not in general definite in the state $|s\rangle$. 
While the results of a measurement of $A$ cannot be predicted for any individual system,
we now show that the frequencies of the results {\it can} be predicted for infinite ensembles
of identically prepared systems.  This is done by demonstrating that the {\it frequency of
occurrence of a given value $a_k$ is a definite quantity for the infinite ensemble}, 
itself considered as an individual system, and further that the definite value in the state
defined in Eq.~\eqref{three} is $|\langle k|s\rangle|^2$.  In this way, the statistical
predictions of quantum mechanics will be recovered from the nonstatistical assertions
about individual systems. 

To carry out this demonstration, the operator $f^k$ for the frequency of occurrence of
the value $a_k$ in an infinite ensemble of individual systems must be constructed.
The state vector of Eq.~\eqref{three} must then be shown to be an eigenvector of $f^k$
with eigenvalue $|\langle k |s\rangle |^2$
\begin{equation}
f^k | (s)^\infty \rangle = |\langle k| s\rangle |^2|(s)^\infty\rangle\, .
\label{four}
\end{equation}
The operator $f^k$ is defined by first constructing the frequency operator $f_{N}^k$ for finite
ensembles of $N$ systems and then taking the limit as $N\to\infty$. The operator
$f_{N}^k$ is easily defined on the Hilbert space for N individual systems in the 
representation in which $A$ is diagonal
\begin{eqnarray}
f_{N}^k & = & \sum_{i_1\cdots i_N}|i_1, 1\rangle | i_2, 2\rangle\cdots|i_N, N\rangle 
\nonumber\\
& \times & \left(N^{-1} \sum^N_{\alpha=1} \delta_{ki_\alpha}\right)\, \langle i_N,
N|\langle i_{N-1}, N-1|\cdots \langle i_1, 1|\, .
\label{five}
\end{eqnarray}
The number in brackets is just the fraction of the states $| i_1\rangle \cdots |
i_N\rangle$  which are the state $k$. As applied to $|(s)^\infty\rangle$ we define
$f_{N}^k$ to be 
\begin{equation}
f_{N}^k |(s)^\infty\rangle  = (f_{N^k} | (s)^N\rangle )  \otimes \ |s, N+1\rangle
 \otimes  |s, N+2\rangle \otimes \cdots\, ,
\label{six}
\end{equation} 
where the term in brackets is defined through Eqs.~\eqref{five} and
\eqref{two}.\footnote{We do not distinguish notationally $f_{N^k}$ as defined on the
Hilbert space of $N$ ensenbles and as defined on the Hilbert space of infinite
ensembles.} The operator $f^k$ acting on $|(s)^\infty\rangle$ is then defined by
\begin{equation}
f^k | (s)^\infty\rangle = \lim_{N\to\infty} f_{N}^k |(s)^\infty\rangle\, .
\label{seven}
\end{equation}
The mathematical meaning and existence of this limit are discussed in the Appendix. To
show that $|(s)^\infty\rangle$ is an eigenfunction�on of $f^k$ with eigenvalue $|\langle
k|s\rangle|^2$ is equivalent to a demonstration of  
\begin{equation}
\| f^k|(s)^\infty\rangle  -  |\langle k|s\rangle|^2| (s)^\infty\rangle||
 =  \lim_{N\to\infty} \| f_{N}^k | (s)^\infty\rangle  -  |\langle k|s\rangle
|^2|(s)^\infty\rangle \|=0\, ,
\label{eight}
\end{equation}
where for any vector $|V\rangle, \|\, |V\rangle \|^2= \langle V|V\rangle$. In computing
the norm in the second term of Eq.~\eqref{eight}, one sees from the definition of
$f_{N^k}$ acting on $|(s)^\infty\rangle$ [Eq.~\eqref{six}] that the normalized states of
systems $N+1, N+2, \cdots$ contribute only factors of unity to the norm so that 
\begin{equation}
\| f^k | (s)^\infty\rangle  -  | \langle k|s\rangle |^2| (s)^\infty\rangle||
= \lim_{N\to\infty} \| f_{N}^k | (s)^N\rangle  -  |\langle k|s\rangle |^2 |
(s)^N\rangle \|_N \, .
\label{nine}
\end{equation}
Here, we use a subscript $N$ to indicate that the norm is to be taken in the Hilbert
space for ensembles of $N$ identical systems.

To show that the limit in Eq.~\eqref{nine} vanishes, we write out the norm using the
definition Eq.~\eqref{five} and the orthonormality of the states $|i\rangle$ to find
%\begin{eqnarray}
%\sum_{i_1\cdots i_N} \Biggl[\sum^N_{\alpha, \beta=1} \left(N^{-2}\delta_{ki_\alpha}
%\delta_{ki_\beta}\right) & - & 2 |\langle k|s\rangle |^2 \sum^N_{\alpha=1} \left(N^{-1}
%\delta_{ki_\alpha}\right)\nonumber\\
%+ | \langle k|s\rangle|^4\Biggr]\, |\langle i_1 | s\rangle |^2 & \cdots & |\langle
%i_N|s\rangle |^2\, .
%\label{ten}
%\end{eqnarray}
\begin{equation}
\sum_{i_1\cdots i_N} \Biggl[\sum^N_{\alpha, \beta=1} \left(N^{-2}\delta_{ki_\alpha}
\delta_{ki_\beta}\right)  -  2 |\langle k|s\rangle |^2 \sum^N_{\alpha=1} \left(N^{-1}
\delta_{ki_\alpha}\right)
+ | \langle k|s\rangle|^4\Biggr]\, |\langle i_1 | s\rangle |^2  \cdots  |\langle
i_N|s\rangle |^2\, .
\label{ten}
\end{equation}
The sum over $i_1\cdots i_N$ may be done first using the completeness of the states
$|i\rangle$ and the normalization of $|s\rangle$. In doing this for the first term, one
arrives at a different answer if $\alpha=\beta$ than if $\alpha\not=\beta$. The remaining
summands are independent of $\alpha$ and $\beta$ so that the sums over these quantities
are easily performed. The result is
\begin{equation}
\| f^k |(s)^\infty\rangle  -  |\langle k|s\rangle |^2 | (s)^\infty\rangle
\|^2 
 =  \lim_{N\to\infty} \Bigl( (1/N) | \langle k|s\rangle|^2  \\
 +  \left\{[N(N-1)/N^2]-1\right\}  |  \langle k|s\rangle|^4\Bigr)=0\, .
\label{eleven}
\end{equation}

%\begin{align}
%\| f^k |(s)^\infty\rangle & - & |\langle k|s\rangle |^2 | (s)^\infty\rangle
%\%|^2\nonumber\\
%& =&  \lim_{N\to\infty} \Bigl( (1/N) | \langle k|s\rangle|^2
%& + & \left\{[N(N-1)/N^2]-1\right\}  |  \langle k|s\rangle|^4\Bigr)=0\, .
%\label{eleven}
%\end{align}
This completes the demonstration. For an infinitely large ensemble the frequency of
occurrence of a value $\alpha_k$ is a definite quantity with value $|\langle
k|s\rangle|^2$.

Not all of the statistical predictions of quantum mechanics are for ensembles of
individual systems, but other statistical assertions can be derived from these. By way of
example, consider repeated measurements of a quantity $A$ on a single system separated by
a time interval $\Delta t$.  Quantum mechanics predicts that the probability of a result
$a_n$, succeeding a result $\alpha_m$, is $|U_{nm}|^2$, where $U_{nm}=\langle m|\exp
(-iH\Delta t)|n\rangle$ and $H$ is the Hamiltonian for the system. Translated into
empirical terms this is an assertion that if in the sequence of measurements of $A$ the
value $a_m$ occurs $M$ times, the number of succeeding measurements which take the
value $a_n$ will be $M|U_{nm}|^2$ in the limit as $M$ arbitrarily large.

A probability assertion of the above type can be identified with an assertion about an 
ensemble of systems in the following manner: Imagine an apparatus which records the
results of the measurements. After every measurement which yields the value $a_m$ it
records the value of the succeeding measurement. If, at the end of the sequence of 
measurements the value $a_m$ occurred $M$ times, it has $M$ variables which indicate the
value of the succeeding measurement. If we denote the state of this apparatus in which
these $M$ variables have the value $a_{n_1}\cdots a_{n_M}$ by $| n_1, \cdots
n_M\rangle$ (where the remaining variables have been suppressed), then the laws of
quantum mechanics tell us that the state vector for the apparatus at the end of 
the sequence of measurements is
\begin{equation}
\sum_{n_1\cdots n_M} U_{n_1m} U_{n_2m} \cdots U_{n_Mm}|n_1\cdots n_M\rangle\, .
\label{twelve}
\end{equation}
This state vector, however, has the same properties as that of Eq.~\eqref{two} , 
and the same argument can be adduced to show that the frequency of occurrence of the value $n$
among the $n_1\cdots n_M$ is definite and given by $|U_{nm}|^2$ in the limit of infinite $M$.
In this way, this type of statistical prediction of quantum mechanics can also be deduced from
the quantum mechanics of individual systems.

The probability predictions of quantum mechanics, interpreted as predictions of the frequencies
of results of measurements on infinite ensembles of identically prepared systems, are thus seen
not to enter into the theory in any preferred way, but have the same status as any other
observable in the theory.\footnote{Substantially the same analysis and conclusions as are given
in this section have been obtained independently by R.N.~Graham in a University of North
Carolina doctoral dissertation, preliminary versions of which were prepared in the summer of
1967. The derivation of the statistical predictions of quantum mechanics from its assertions
about individual systems has also been considered by H.~Everett, {\sl Rev.~Mod.~Phys.} {\bf 29}, 
454 (1957), but from a rather different point of view than that given here. Both Everett's theory
and Graham's work have been reviewed by B.~DeWitt in {\sl Battelle Recentres} (W.A.~Benjamin, Inc.,
1968), Appreciation is expressed to B.~DeWitt for helpful discussion and correspondence on these
points.} 

\section{Is the State an Objective Property of a System?}

In Sec.~I, a definition of state was given for classical and quantum-mechanical systems. We
now turn to the question of whether the state so defined can be regarded as an objective
property of the system. It is just this point which seems central in many of the difficult
problems of interpretation of quantum mechanics.

The state of the system will be called an objective property of the system if it can be
determined by measurements on the system in complete ignorance of its previous condition. More
precisely, the {\sl state will be called an objective property if an assertion of what the state is
can be verified by measurements on the individual system without knowledge of the system's
previous history}.

In classical physics the state is an objective property of the physical system. For example,
for a single particle the state may be characterized by the position {\bf x} and the momentum
{\bf p}. To determine the state one has only to measure {\bf x} and {\bf p}. Further,
specification of the state is the same as specifying the numbers {\bf x} and {\bf p} and can 
be verified by measuring these quantities. Generally, a classical state is an objective property 
because it lists for {\it all} propositions the truth values, {\it true} or {\it false}, and can, 
therefore, be verified by comparing this list with one obtained by measurements on the individual 
system. The situation is quite different in quantum mechanics.

In quantum mechanics there are no measurements on an individual system which can determine its
state. There are no measurements on an individual system by which an assertion that the state is
described by a certain vector can be verified. For example, if a single particle is asserted to 
be in a momentum eigenstate, a measurement of the momentum which yields the eigenvalue does not
distinguish between this state and one which the momentum is indefinite. A quantum-mechanical
state is, therefore, not an objective property of the individual system.

The fact that a quantum-mechanical state is not an objective property of an individual system
does not imply that it cannot be known. If this were so, the idea of state would not be a very
useful one. A state is known when it is determined how the system has been prepared. The
assertion that systems prepared in a certain way have a characteristic state can be verified by
measurements on many systems prepared in the same way. In the same way the wave function of a
single system (or at least its absolute square) can be determined by measurements of an
ensemble of systems {\it provided that} it is known that they have been identically prepared. In the
absence of such knowledge, the wave function cannot be determined from the frequencies of the
various measurement results. This is in contrast to the situation in classical physics where
the state can be determined without any information about the past history of the system. In
quantum mechanics, because the state cannot be determined without some knowledge of how the
system was prepared, the state is not an objective property of the system. 

The fact revealed by the preceding analysis should already have been clear from the definition
of quantum-mechanical state given in Sec. I. A state is a list of all measurable propositions
together with the values {\it true}, {\it false}, or {\it indefinite}. The property of a
proposition being indefinite can hardly be an objective property of the physical system, but
rather is a statement of the knowledge of the observer about this proposition. The state is not
an objective property of an individual system but is that information, obtained from a
knowledge of how the system was prepared, which can be used for making predictions about future
measurements.

The state of a classical system is an objective property of the system and therefore changes
only by dynamical laws. A quantum-mechanical state being a summary of the observers'
information about an individual physical system changes both by dynamical laws, and whenever
the observer acquires new information about the system through the process of measurement. The
existence of two laws for the evolution of the state vector by the Schr\"odinger equation on the
one hand and by the process of measurement (sometimes described as the ``reduction of the wave
packet'') on the other, is a classic subject for discussion in the quantum theory of
measurement. The situation becomes problematical only if it is believed that the state vector
is an objective property of the system. Then, the state vector must be required to change only
by dynamical law, and the problem must be faced of justifying the second mode of evolution from
the first. If, however, the state of a system is defined as a list of propositions together
with their truth values, it is not surprising that after a measurement the state must be
changed to be in accord with the new information (if any) acquired about the state after the
measurement and with the old information lost because of the irreducible disturbance of the
system caused by the measurement. The ``reduction of the wave packet'' does take place in the
consciousness of the observer, not because of any unique physical process which takes place
there, but only because the state is a construct of the observer and not an objective property
of the physical system.

In conclusion, we review how the understanding of ``state of a system'' resolves some of the
so-called ``paradoxes'' of quantum mechanics. The first we consider is one raised in the famous
paper of Einstein, Podolsky, and Rosen\footnote{A.~Einstein, B.~Podlsky, and N.~Rosen, {\sl Phys.
Rev.} {\bf 47}, 777 (1935).} phrased in a slightly different language. We have two spin-$\frac{1}{2}$
particles known to be in a state of total angular momentum zero. They scatter and separate. The
state vector for the combined system is
\begin{equation}
\alpha |\, \uparrow\, , 1\rangle |\, \downarrow\, , 2\rangle + \beta |\, \downarrow\, , 1\rangle |
\, \uparrow\, , 2\rangle\, ,
\label{thirteen}
\end{equation}
where $|\, \uparrow\, i\rangle$ and $|\, \downarrow\, , i\rangle$ are the states in which the spin
of particle $i$ is parallel or antiparallel to some determined axis. In this state no definite
prediction can be made about the spin of either particle, but only probability predictions for
ensembles of identical scatterings. If a measurement of the spin of particle one obtains the value
$\uparrow\, ,$ then the laws of quantum mechanics assert that the state immediately changes to
\begin{equation}
|\, \uparrow\, , 1\rangle\, |\, \downarrow\, , 2\rangle\, .
\label{fourteen}
\end{equation}
The spin of the second particle can now be predicted with certainty to be $\downarrow$. 

The paradox arises by asking how a measurement on the first particle can change the description of
the state of the second (which might be very far away from the first) from one in which the spin
is indefinite to one in which the spin is definitely $\downarrow$ . This can be a paradox, however, 
only if the
state is regarded as an objective property of the physical system which can change suddenly only
by the result of a sudden interaction with another system. As regarded here, however, a
description of a state is a description of the information possessed by the observer about the
system. A measurement of the spin of particle two because of the special way the spins are
correlated in this experiment. This new information is summarized by associating the vector 
$|\, \downarrow\, , 2\rangle$ with the second particle.

The second paradox we consider has been discussed by Wigner.\footnote{See Ref.~1h.} Two friends
are participating in an experiment. The first friend makes a measurement on an object which, let
us say, has two possible results, $a$ and $b$. The second friend, sometime after the measurement, asks
the first about the result and obtains an appropriate reply. Let us suppose the vectors
$|a\rangle$ and $|b\rangle$ describe the state of the object in which the results of the experiment 
are $a$ and $b$, respectively, and that the vectors $| A\rangle$ and $| B\rangle$
 describe the state of the first friend in which
he gives, respectively, the reply $a$ or $b$ to the question of the second. If the state of the object
before the measurement should be summarized by the vector 
\begin{equation}
\alpha | a\rangle + \beta | b\rangle\, ,
\label{fifteen}
\end{equation}
then the second friend will describe the combined system of object plus first friend by the vector 
\begin{equation}
\alpha | a\rangle\, |A\rangle + \beta\, |b\rangle\, |B\rangle\, .
\label{sixteen}
\end{equation}
This will be the correct description from the time of the measurement until the second friend
learns from the first the result of the measurement at which point the state vector is changed to
either $|\alpha\rangle |A\rangle$ or $|b\rangle | B\rangle$.  If, however, the first friend is
asked how he described the object between the time of the measurement and the time he answers the
question, he will reply with either the vector $|a\rangle$ or $|b\rangle$. In his mind the result
of the measurement is already determined. If one believes that the state vector is an objective
property of the object, one is forced to conclude that the vector, which describes the system of
object and first friend after the measurement, is then either $|a\rangle |A\rangle$ or $|b\rangle
|B\rangle$ and not that of Eq.~\eqref{sixteen} thus violating the linear laws of quantum
mechanics. The state is, however, not an objective property of the system but a convenient
shorthand for the information assertible by each friend. It is not inconsistent for the first
friend to describe the object by the state vectors $|a\rangle$ or $|b\rangle$ at the same time as
the second friend describes the combined system by the vector of Eq.~\eqref{sixteen}, 
because each state vector represents correctly the information each friend has as a consequence of
his knowledge of how the state of the object was prepared and (in the case of the first friend)
the result of his subsequent measurements.

\acknowledgments

Appreciation is expressed to B.~DeWitt, A.~Dragt, M.~Rosenfeld, H.~Shepard, J.R.~Taylor, and 
C.H.~Woo for helpful conversations. Special thanks are due to E.P.~Wigner for several critical
discussions.

\section*{Mathematical Appendix}

The argument used in Sec. III to show that the frequency distribution of any observable is a
definite quantity in an infinite ensemble of identically prepared systems, is here given in a
mathematically precise form. To accomplish this the construction of the Hilbert space for the
infinite ensemble is first briefly reviewed following the work of von Neumann.\footnote{J.~von
Neumann, {\sl Compositio Math.} {\bf 6}, 1, (1939). For a more recent treatment see M.A.~Guichardet, 
{\sl Ann.~Sc.~Ec.~Norm.~Sup. 3e Serie} {\bf 83}, 1 (1966).} The operator $f_{N^k}$ for a finite 
ensemble 
is then defined on this Hilbert space and the existence of the limit $f^k$ as $N\to \infty$
proved. The state vector representing an infinite ensemble of identically prepared systems each in
state $|s\rangle$ is then shown to be an eigenfunction of this operator with eigenvalue $|\langle
k |s\rangle|^2$.

The Hilbert space for an infinite ensemble of systems is the infinite tensor product ${\cal
H}_1\otimes\, {\cal H}_2\otimes \cdots $ of the Hilbert spaces ${\cal H}_\alpha$ of the individual
physical systems. This infinite tensor product is denoted by ${\cal H}^\infty$ and is precisely
defined in the following. For proof of any of the assertions used in the construction of ${\cal
H}^\infty$ below, the reader is referred to the work of von Neumann, which we follow.

From the Cartesian product ${\cal H}_1 \times {\cal H}_2\times \cdots$ of the individual Hilbert
spaces distinguish all those sequences $|s_1, 1\rangle \otimes |s_2, 2\rangle \otimes |s_3,
3\rangle \otimes \cdots, |s_\alpha \rangle \in {\cal H}_\alpha$ for which the product of the
length of the individual vectors,
\[
\prod^\infty_{\alpha=1} | \langle s_\alpha, \alpha |s_\alpha, \alpha\rangle |^{\frac{1}{2}}\, ,
\]
converges or diverges to zero in the usual sense of infinite products. The class of all such
sequences is denoted by ${\cal C}$.~A scalar product between any two elements of ${\cal C}$,
$|S\rangle = |s_1, 1\rangle \otimes |s_2, 2\rangle \otimes \cdots$ and $|S^\prime \rangle =
|s^\prime_1, 1\rangle \otimes |s^\prime_2, 2\rangle \otimes \cdots$ may  be defined by
\[
\langle S|S^\prime\rangle = \prod^\infty_{\alpha=1} \langle s_\alpha, \alpha | s^\prime_\alpha,
\alpha\rangle\, ,
\]
if this product converges, and by 
\[
\langle S|S^\prime\rangle = 0\, ,
\]
if it does not. Consider now the space ${\cal C}^\prime$ of all finite linear combinations of the
elements of ${\cal C}$ with
\[
  \ |S\rangle = \sum^P_{q=1} \lambda_p | S_p\rangle, \quad   | S_p\rangle \in {\cal C}
\]
and
\[
 |S^\prime\rangle = \sum^Q_{q=1} \mu_q |S^\prime_q\rangle, \quad   | S^\prime_q \rangle \in {\cal C}\, 
\]

Then a scalar product may be defined on ${\cal C}^\prime$ by
\[
\langle S|S^\prime\rangle = \sum^P_{p=1} \sum^Q_{q=1} \lambda_p^* \mu_q \langle S_p |
S^\prime_q\rangle\, .
\]
The space ${\cal C}^\prime$ is thus a linear vector space with a Hermitian scalar product defined.
It is not a Hilbert space, however, because it is not complete; {\it i.e.} any sequence
$|S_n\rangle$ with the property
\[
\lim_{n, m\to\infty} \|\, |S_n\rangle - | S_m\rangle \| =0
\]
(a Cauchy sequence) does not necessarily define a unique limit $|S\rangle$ in ${\cal C}^\prime$
such that
\[
\lim_{n\to\infty} \|\, |S \rangle - |S_n\rangle\, \|=0\, .
\]
While the space ${\cal C}^\prime$ is not complete, it can be completed in the standard manner\footnote{See 
for example, A.E.~Taylor, {\sl Introduction to Functional Analysis} (John Wiley \& Sons, Inc., New York, 1958), 
Secs.~2.41 and 3.21.}  by adjoining all properly identified Cauchy sequences or by embedding ${\cal C}^\prime$ 
in the space of all antilinear functionals on ${\cal C}^\prime$ and then adjoining the limit points in this
space as described by von Neumann. Neither of
these procedures will be outlined here. The resulting complete linear space is the Hilbert space
${\cal H}^\infty$. This Hilbert space is not separable. The space ${\cal C}^\prime$ is dense in
${\cal H}^\infty$ so that for every element $|S\rangle \in {\cal H}^\infty$ there is a sequence of
elements $|S_n\rangle \in {\cal C}^\prime$ with
\[
\lim_{n\to\infty} |S_n\rangle = |S\rangle\, .
\]
The scalar product in ${\cal H}^\infty$ is then given in terms of the scalar product in ${\cal
C}^\prime$ by
\[
\langle S|S^\prime\rangle = \lim_{n\to\infty} \langle S_n | S^\prime_n\rangle\, ,
\]
where
\[
|S^\prime\rangle = \lim_{n\to\infty} |S^\prime_n\rangle\, .
\]
Having constructed the Hilbert space ${\cal H}^\infty$ we begin the proof of the existence of
$f^k$ by defining a subspace ${\cal J}$ of ${\cal H}^\infty$ which contains the state vectors of
all infinite ensembles of identically prepared systems. Indeed, we define ${\cal J}$ to be the 
smallest subspace of ${\cal H}^\infty$  containing all vectors of the form
\[
|(s)^\infty\rangle = |s, 1\rangle \otimes\, |s, 2\rangle \otimes\, |s, 3\rangle \otimes \cdots, \quad
\langle s|s\rangle=1\, .
\]
Clearly, $|(s)^\infty\rangle$ is contained in ${\cal H}^\infty$ since it is contained in ${\cal
C}$.  ${\cal J}$ is then the set of all finite linear combinations of vectors of this form
together with the limits of their Cauchy sequences. The limits exist because ${\cal H}^\infty$ 
is complete and ${\cal J}$ is then a subspace.

The frequency operator $f_{N^k}$ for finite ensembles was defined on vectors of the form
$|(s)^\infty\rangle$ by Eqs.~\eqref{five} and \eqref{six}, and by linearity it can be extended
to all finite linear combinations of vectors of this form. The norm of $f_{N^k}$  acting on
$|(s)^\infty\rangle$ is given by a calculation similar to that of Sec.~III 
\begin{eqnarray*}
\| f_{N^k} |(s)^\infty \rangle \|^2 & = & \| f_{N^k} |(s)^N\rangle \|^2_N \\
= (1/N)\ \bigl[\, |\langle k|s\rangle |^2 & + & (N-1)\, |\langle k | s\rangle |^4\bigr]\\
\leq  1 & = & \|\, |(s)^\infty\rangle\, \|^2\, .
\end{eqnarray*}
Now, it is easily verified that $f_{N^k} |(s)^\infty\rangle$ is orthogonal to $f_{M^k} |
(s^\prime)^\infty\rangle$ if $|s\rangle \not= |s^\prime\rangle$, for if we take $N\geq M$, then
\begin{equation*}
\bigl\langle (s^\prime)^\infty | f_{N^k} f_{M^k} | (s)^\infty\bigr\rangle  =  \left\langle
(s^\prime)^N |f_{N^k} f_{M^k} | (s)^N\right\rangle_N
  \prod^\infty_{\alpha=N+1} \langle s|s^\prime\rangle=0
\end{equation*}
as the last product always diverges to zero. It then follows immediately that if $|S\rangle$ is
any finite linear combination of vectors of the form $|(s)^\infty\rangle$ that
\[
\| f_{N^k} |S\rangle \, \|\leq \|\, |S\rangle \|\, .
\]
Since the set of all finite linear combinations is dense in ${\cal J}$, it follows that
$f_{N^k}$ can be extended to the whole of ${\cal J}$ by continuity and satisfies the same
bound. The operator $f_{N^k}$ is conveniently extended to the whole Hilbert space ${\cal
H}^\infty$ by defining
\[
f_{N^k} | S\rangle = 0, \quad |S\rangle \notin {\cal J}\, .
\]
Having defined the operator $f_{N^k}$ on the whole Hilbert space ${\cal H}^\infty$, we now
consider the limit as $N\to \infty$. To show the existence of this limit we first note that for
$N\geq M$
\begin{eqnarray*}
\| f_{N^k} | (s)^\infty\rangle & - & f_{M^k} | (s)^\infty \rangle \|^2\\
& = & \| f_{N^k} | (s)^N\rangle  -  f_{M^k} | (s)^N \rangle||^2_N\\
& = & (M^{-1}-N^{-1}\, \bigl[\, |\langle k|s\rangle\, |^2 - | k|s\rangle |^4\bigr]\\
& \leq & (M^{-1}- N^{-1})\, \|\, |(s)^\infty\rangle\, \|^2
\end{eqnarray*}
where the calculation is made using the same techniques as in Sec. III. If use is made of the
orthogonality between $f_{N^k} | (s)^\infty\rangle$ and $f_{M^k} | (s^\prime)^\infty\rangle$, 
the bound may be extended to all finite linear combinations of vectors of the form
$|(s)^\infty\rangle$ and by continuity (since $f_{N^k}-f_{M^k}$ is bounded) to the whole of
${\cal J}$. The bound also clearly holds for vectors not in ${\cal J}$, so we have it on the
whole Hilbert space. Thus, we have for any $| S\rangle\, \in {\cal H}^\infty$
\begin{equation*}
\lim_{M, N\to\infty} \| f_{N^k} | S\rangle  -  f_{M^k} | S\rangle \|\\
\leq \lim_{M, N\to\infty} (M^{-1}  -  N^{-1})^{\frac{1}{2}}\|\, |S\rangle\|=0\, .
\end{equation*}
%%\begin{equation}
%\lim_{M, N\to\infty} \| f_{N^k} | S\rangle  -  f_{M^k} | S\rangle \||
%\leq \lim_{M, N\to\infty} (M^{-1}  - |N^{-1})^{\frac{1}{2}}\|\, |S\rangle\|=0\, .
%\end{equation}
The sequence of operators $f_{N^k}$ thus converges strongly to an operator $f^k$ defined on the
complete space ${\cal H}^\infty$ by
\[
f^k|S\rangle = \lim_{N\to\infty} f_{N^k} |S\rangle\, .
\]
In particular, the calculation given in Sec. III now goes over without reservation to show
\begin{equation*}
\| f^k |(s)^\infty\rangle  - | \langle k|s\rangle |^2\, |(s)^\infty\rangle\, \|^2
= \lim_{N\to\infty} \|f_{N^k} |(s)^\infty\rangle  -  |\langle k|s\rangle |^2
|(s)^\infty\rangle\, \|^2=0 .
\end{equation*}
which implies
\[
f^k|(s)^\infty\rangle = |\langle k|s\rangle\, |^2\, |(s)^\infty\rangle\, .
\]

\end{document}